\documentclass[pre,twocolumn]{revtex4}
\usepackage[english]{babel}
\usepackage{amsmath}
\usepackage{amssymb}
\usepackage{epsfig}
\usepackage{url}

\begin{document}
\title{Stabilization of Optical Bubbles Near the Axis of a Helical Waveguide}
\author{Victor P. Ruban}
\email{ruban@itp.ac.ru}
\affiliation{Landau Institute for Theoretical Physics RAS,
Chernogolovka, Moscow region, 142432 Russia}

\date{\today}

\begin{abstract}
It has been shown numerically that coupled nonlinear Schrödinger equations
describing the interaction between the left and right circular polarizations
of a paraxial optical wave in a defocusing Kerr medium with an anomalous
dispersion in a helical waveguide have stable solutions in the form of
elongated stationary rotating bubbles with several optical vortices attached
to the ends. A bubble is an arbitrarily long quasi-cylindrical three-dimensional
cavity in one of the components filled with the opposite component. 
The transverse profile of the bubble is determined by the shape of the cross
section of the waveguide, the helix pitch, the number of vortices, and the
background intensity of the surrounding component rather than by the total
amount of the filling component.

\vspace{5mm}

\noindent DOI: 10.1134/S0021364024602264
\end{abstract}

\maketitle

\subsection*{Introduction}

When studying nonlinear waves, long-lived coherent 
structures, in particular, solitons and vortices, are
of great interest (see, e.g., [1--16], and references
therein). New types of such objects are obtained in
laboratory and numerical experiments. Multicomponent 
wave systems are particularly rich in this context
and optical systems are the most obvious examples
because light can have two independent polarizations.
Consequently, it is reasonable to consider an optical
wave under ``unconventional'' conditions in order to
possibly detect new effects. The success of this
approach is confirmed in this work, where it is shown
that the nonlinear interaction between polarizations
can form previously unknown wave structures under
the action of a special spatial inhomogeneity.

A weakly nonlinear quasi-monochromatic wave propagating
paraxially in a wide helical waveguide was chosen for 
this study because of the formal analogy between nonlinear
optics and diluted Bose--Einstein condensates. As known, 
the paraxial propagation of light with two circular
polarizations in a locally isotropic Kerr medium is 
described by two coupled nonlinear Schrödinger equations [17]. 
In the case of defocusing nonlinearity and anomalous dispersion,
equations in form coincide with Gross--Pitaevskii equations
for a binary Bose--Einstein condensate of cold atoms in
the phase separation regime [18--24]. A two-component 
system allows the existence of domain walls separating 
regions with right and left circular polarization [25--31]. 
The combination of domain walls and quantized 
vortices provides interesting three-dimensional 
structures (see [32--36] and references therein).
Stable vortex structures in Bose--Einstein condensates
are maintained by the rotation of the trap potential. In
optical systems, the (rescaled) coordinate $\zeta$ along the
light propagation direction rather than the time $t$
serves as the evolution variable, and the helical
symmetry of the waveguide is analog of rotation [37],
when the permittivity profile depends on two combinations 
of the variables:
$$
\tilde x= x \cos\Omega \zeta +y \sin\Omega \zeta, \qquad
\tilde y= y \cos\Omega \zeta -x \sin\Omega \zeta.
$$
In this case, the helix pitch $S=2\pi/\Omega$ is an analog of 
the time period of trap rotation. The ``retarded'' time
$\tau=t-\zeta/v_{\rm gr}$ serves as the third ``spatial'' 
coordinate in optical systems. A significant practical difference
between Bose--Einstein condensates and light beams
is the translational symmetry along this third coordinate
in the latter case. Therefore, longitudinally delocalized 
solutions of equations are not relevant for the physics of
cold gases but are directly applicable for optical systems.

Some quasistationary configurations of binary optical 
beams, where both components of the light wave are 
nearly ``equivalent,'' have been numerically obtained 
in recent work [37]. In this work, stable stationary
solutions of a new type --- arbitrarily long light
``bubbles'' (bunches of the conditionally second component 
inside the conditionally first component),
which are confined near the axis of the helical waveguide 
by several quantized vortices --- are obtained. It should
be emphasized that the existence of at least two vortices 
in the first component is necessary. As shown in [35],
one vortex cannot confine a large bubble near the axis. 
The main properties of new three-dimensional 
structures are theoretically interpreted through
the analysis of the energy dependences for strictly 
two-dimensional solutions with the content $n$ of the second
``light fluid'' in the presence of $Q$ vortices. The formation 
of quasi-one-dimensional bubbles is similar to the
formation of a condensed phase adjacent to vacuum for an 
imaginary substance with the energy density $\varepsilon(n)$
if the pressure $p=n^2d(\varepsilon(n)/n)/dn$ vanishes at a
certain finite density $n_0$. In other words, the function
$\varepsilon(n)/n$ has a nontrivial minimum (at the point where
the $\varepsilon(n)$ plot touches a straight line passing through the
coordinate origin). The value $n_0$ strongly depends on
the ``rotation velocity'' $\Omega$, the number of vortices $Q$,
and the corresponding ``total density'' $n_1+n$ (see, e.g.,
plots in Fig. 1). The length of the condensed phase
segment can be arbitrary.

\begin{figure}
\begin{center} 
\epsfig{file=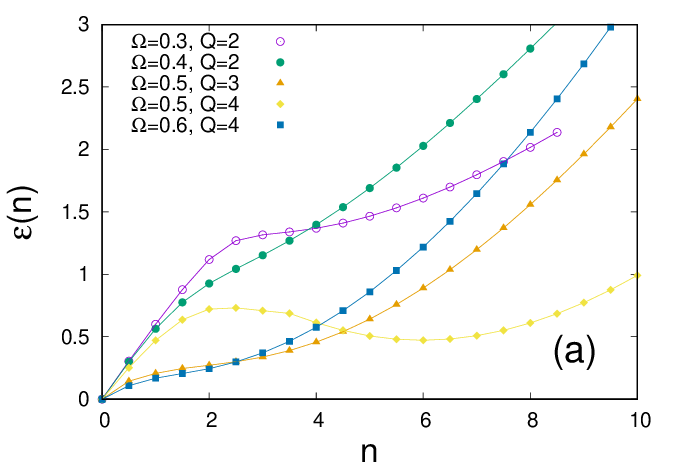, width=84mm}\\
\epsfig{file=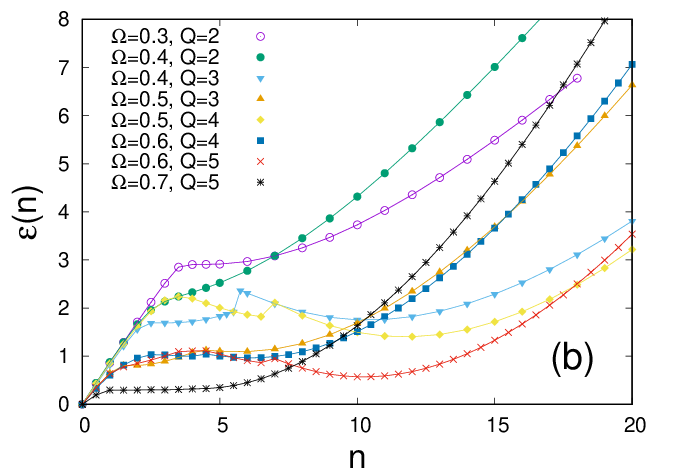, width=84mm}
\end{center}
\caption{
Energy dependences for various rotation velocities and the number of 
vortices at the total density $n_+=(n_1+n)=$  (a) 60.0 and (b) 100.0.
}
\label{Energy_2D} 
\end{figure}

\subsection*{Equations and the numerical method}

As in previous works [35, 37], we consider a transparent optical 
medium with a defocusing Kerr nonlinearity and with the dispersion 
relation $k(\omega)=\sqrt{\varepsilon(\omega)}\omega/c$
for linear waves and assume the existence of
a frequency interval with anomalous dispersion
$k''(\omega)<0$. It is near the low-frequency edge of the
transparency window (often the infrared spectral band
in real materials; see, e.g., [38, 39]). In this situation,
it is possible to apply the known equation for the vector 
envelope of a weakly nonlinear quasimonochromatic light wave 
(with the carrier frequency $\omega_0$; the complex exponential 
is chosen in the form $\exp[ik_0\zeta-i\omega_0 t]$) 
in the paraxial approximation:
\begin{eqnarray}
&&2k_0[-i\partial_\zeta-ik_0'\partial_t+k_0''\partial_t^2/2]{\bf E}
-\Delta_\perp {\bf E}\approx\nonumber\\
&&\approx\frac{k_0^2}{\varepsilon(\omega_0)}\Big[
\tilde\varepsilon(x,y,\zeta){\bf E}
+\alpha|{\bf E}|^2{\bf E}+\beta({\bf E}\cdot{\bf E}) {\bf E}^*\Big].
\label{E_eqs}
\end{eqnarray}
Here, $k'_0=1/v_{\rm gr}$ is the inverse group velocity of light in
the medium, $k''_0$ is the negative chromatic dispersion coefficient, 
$\tilde\varepsilon(x,y,\zeta)$ is the small inhomogeneity 
of the permittivity at the carrier frequency, and $\alpha(\omega_0)$ 
and $\beta(\omega_0)$ are the negative nonlinear coefficients.
The retarded time $\tau=t-\zeta/v_{\rm gr}$ is introduced as a new 
variable. The amplitude of the electric field is expressed in
terms of the slow amplitudes $A_{1,2}(x,y,\tau,\zeta)$ of the left
and right circular polarizations as
\begin{equation}
{\bf E}\approx \big[({\bf e}_x+i{\bf e}_y) A_1 
+ ({\bf e}_x-i{\bf e}_y) A_2 \big]/\sqrt{2}. 
\end{equation}
Then, the light wave is described by a pair of coupled nonlinear 
Schrödinger equations [17], similar to a binary Bose--Einstein 
condensate of cold atoms (with the change in the variables 
$\zeta \rightarrow t$, $\tau \rightarrow z$ ). After rescaling, 
the dimensionless system is obtained in the form
\begin{equation}
i\frac{\partial A_{1,2}}{\partial \zeta}=
\Big[-\frac{1}{2}\Delta +U(x,y,\zeta)
+|A_{1,2}|^2 + g|A_{2,1}|^2\Big]A_{1,2},
\label{A_12_eqs}
\end{equation}
where  $\Delta=\partial_x^2+\partial_y^2+\partial_\tau^2$
is the three-dimensional Laplace operator in the ``coordinate'' space 
${\bf r}=(x,y,\tau)$, $U\propto -\tilde\varepsilon(x,y,\zeta)$ 
is the external potential, and $g=1+2\beta/\alpha$ is the cross-phase
modulation parameter, which is equal to 2 under the assumption of fast
nonlinear response.  It is substantial that the nonlinear interaction 
between two components is reduced to a simple incoherent coupling
through the coefficient $g$ and keeps the amount of each component
$N_{1,2}=\int |A_{1,2}|^2 dx dy d\tau$. Thus, the model describes two
quantum compressible liquids with the densities $I_{1,2}=|A_{1,2}|^2$
and velocities ${\bf v}_{1,2}=\nabla \mbox{Arg} (A_{1,2})$.

As in recent work [37], we focus on helical waveguides
with a flat bottom and sharp walls. Such waveguides 
should be easily implemented in experiments, particularly 
for a liquid Kerr medium. However, for the convenient numerical
simulation, the corresponding potential well with the vertical
walls is approximated by the expression
\begin{equation}
U= C[1-\exp(-[(\tilde x^2+\kappa^2\tilde y^2)/36]^5)],
\end{equation}
where $C=42$ and $\kappa^2=(1+0.3)/(1-0.3)=13/7$ is the
transverse anisotropy, which is necessary for the
action of the rotation.

It is fundamentally important that Eqs. (3) constitute 
the Hamiltonian system
$$
i \partial A_{1,2}/\partial \zeta=\delta{\cal H}/\delta A^*_{1,2}.
$$
The corresponding non-autonomous Hamiltonian has the form
\begin{eqnarray}
{\cal H}&=&\frac{1}{2}\int(|\nabla A_1|^2+|\nabla A_2|^2)dx dy d\tau
\nonumber\\
&+&\int U(x,y,\zeta)(|A_1|^2+|A_2|^2)dx dy d\tau \nonumber\\
&+&\frac{1}{2}\int(|A_1|^4+|A_2|^4+2g|A_1|^2|A_2|^2)dx dy d\tau.
\end{eqnarray}
This functional is not conserved in the process of evolution. 
However, since the system in the rotating coordinates is autonomous,
the functional
\begin{equation}
{\cal H}_\Omega={\cal H}
-\Omega\int [A^{\dag} (iy\partial_x-ix\partial_y)A]dx dy d\tau,
\end{equation}
where $A=(A_1,A_2)^T$ is the two-component column, is
an integral of motion.

This work is based on the fact that stationary rotating stable 
solutions of the system of Eqs. (3) are ``local minimum points''
for the functional ${\cal H}_\Omega$ (written in the variables 
$\tilde x$ and $\tilde y$) at fixed $N_{1}$ and $N_{2}$ values.

Equations (3) were numerically simulated using the standard 
split-step Fourier method of the second order in the evolution 
variable $\zeta$ in the initial (nonrotating) coordinate system. 
The computational region in the variables $x$, $y$, and $\tau$ was
a $6\pi\times 6\pi\times 12\pi$ rectangular parallelepiped 
with periodic boundary conditions. 
However, since the potential well is sufficiently deep, 
the functions $A_1$ and $A_2$ rapidly decrease in the
transverse directions to almost zero, so that the effect
of transverse boundaries is negligibly small.

The accuracy of calculations was controlled by the
conservation of integrals of motion to 4--6 decimal
places in the interval $0< \zeta < 600$ (which corresponds
to a propagation length of several tens of meters for a
transverse scale of about several tens of wavelengths).

A nearly steady initial state was numerically prepared
using a dissipative procedure including two steps in
each cycle. The first step corresponded to the purely
gradient dissipative dynamics
$$
-\partial A_{1,2}/\partial \eta=\delta{\cal H}_\Omega/\delta A^*_{1,2}
$$
in a narrow range of the auxiliary variable $\eta$. In the
second step, the functions $A_1(x,y,\tau)$ and $A_2(x,y,\tau)$ are
multiplied by suitable factors $f_1$ and $f_2$, respectively, in order
to keep the given $N_1$ and $N_2$ values, respectively. The
total $\eta$ range was several tens, so that all hard degrees
of freedom were effectively suppressed and the system
was near the minimum ${\cal H}_\Omega$.

\subsection*{Results}

\begin{figure}
\begin{center} 
\epsfig{file=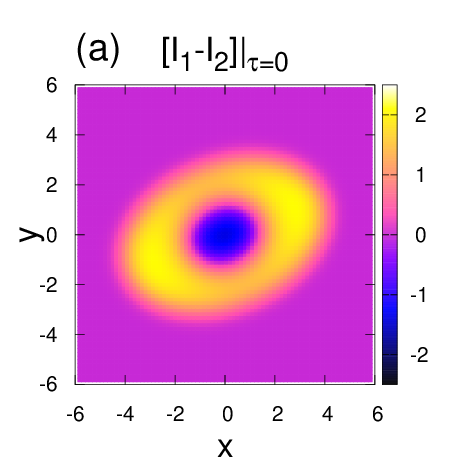, width=42mm}
\epsfig{file=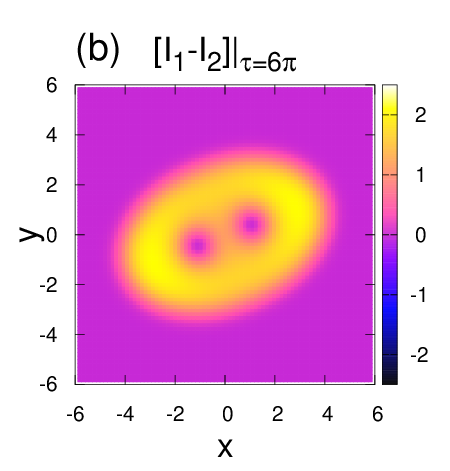, width=42mm}\\
\epsfig{file=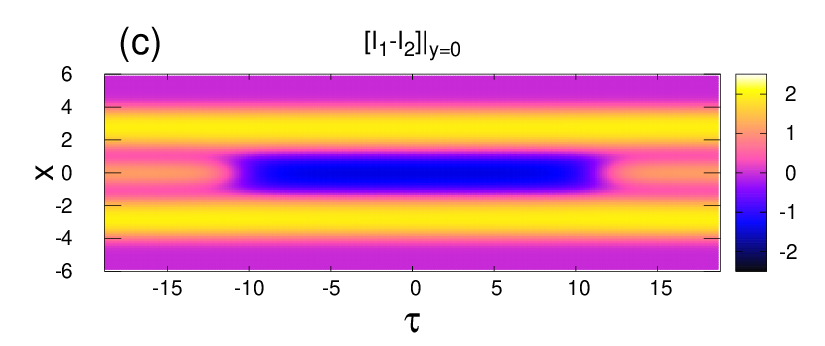, width=86mm}\\
\epsfig{file=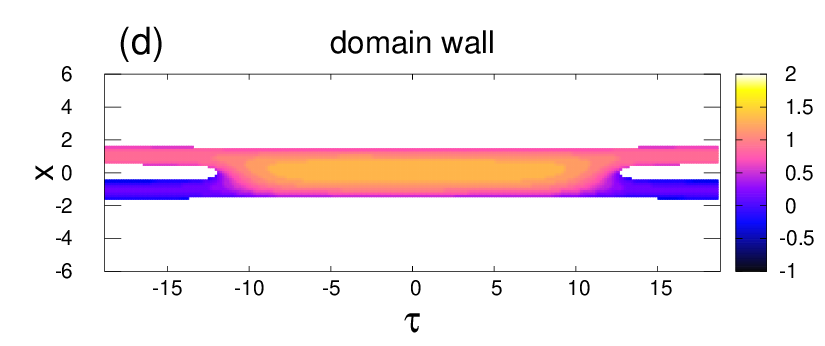, width=86mm}
\end{center}
\caption{
Numerical example of a bubble with two vortices at the 
``rotation frequency'' $\Omega=0.4$ and the filling factor
$\langle n_+\rangle=60$ average in the variable $\tau$ at the
distance $\zeta=590$: (a) cross section of the light beam through
the bubble, (b) cross section through empty vortices,
(c) longitudinal section by the $y=0$ plane, and (d) general
side view of the conditional surface of the bubble and vortex 
filaments attached to it (the color of the points of the
numerical lattice near the middle of the domain wall or in
the core of the vortex corresponds to the $y$ coordinate 
perpendicular to the figure plane).
}
\label{V2B} 
\end{figure}

\begin{figure}
\begin{center} 
\epsfig{file=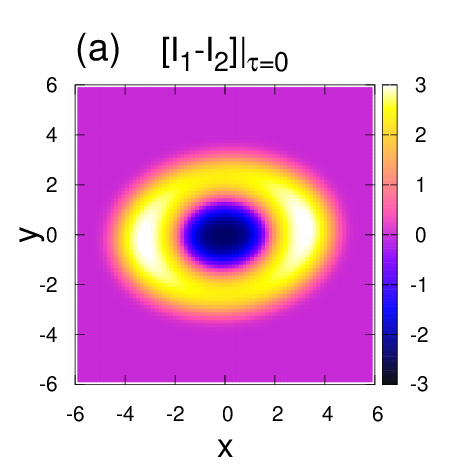, width=42mm}
\epsfig{file=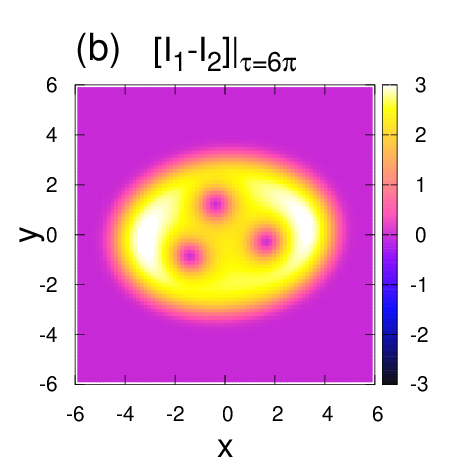, width=42mm}\\
\epsfig{file=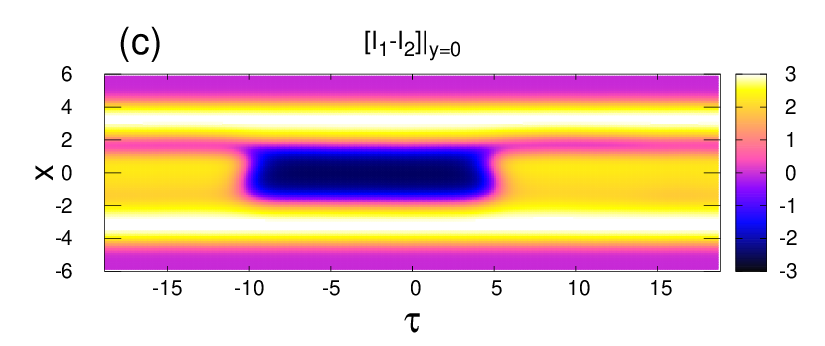, width=86mm}\\
\epsfig{file=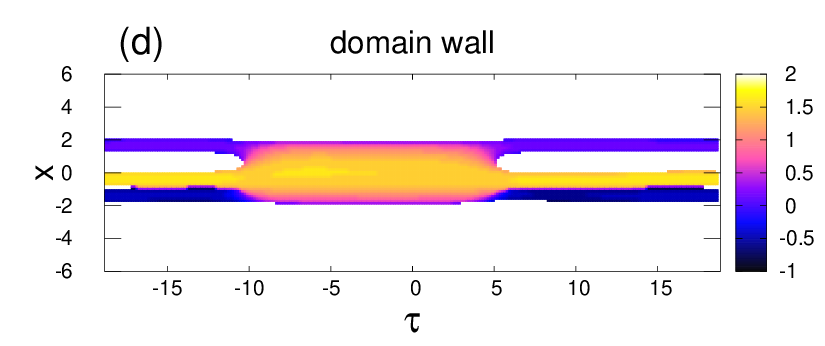, width=86mm}
\end{center}
\caption{
Example of the bubble with three vortices at 
$\Omega=0.4$, $\langle n_+\rangle =100$, and $\zeta=550$.
}
\label{V3B} 
\end{figure}

Preliminary numerical experiments with the initial
conditions in the form of a slightly perturbed filled
multiple vortex were first carried out in the free search
regime. The subsequent dynamics appeared to significantly 
depend on the parameters. Only small oscillations near
the stable longitudinally homogeneous state were observed 
in some cases. However, instability was developed in the system
at low filling; as a result, the second component was redistributed
along the beam axis and characteristic structures in the form of
bubbles with several attached ``empty'' single longitudinal vortices
are formed (they are very similar to those shown in Figs. 2 and 3, 
but less smooth and stationary).

The interpretation of these preliminary experiments 
required a more detailed study of purely two-dimensional 
steady states. To find such states with a good accuracy, 
the typical range of the variable $\eta$ in the
dissipative procedure was several hundreds and sometimes 
reached thousands. Long-term relaxation occurred in the 
regime intermediate in the variable $n=\int I_2 dx dy$ 
between configurations with all almost empty separated vortices 
on one side and single filled multi-charged vortex at 
sufficiently large $n$ values on the other side. 
In this intermediate regime, only some vortices 
are joined in a filled vortex with a smaller multiplicity 
and the other vortices ``prefer'' to be arranged
separately (sometimes asymmetrically). The presence
of several local minima of the energy functional with
their attraction regions can lead to kinks or even jumps
on the corresponding dependences of the energy on
the variable $n$. Fortunately, the aims of this work do
not require the insight into all these difficulties
because the intermediate region is certainly unstable
with respect to three-dimensional perturbations. The
existence of stable solutions in the regions of small $n$ 
and sufficiently large $n$ values is practically important.

Figure 1 presents the dependences of the energy of
two-dimensional solutions on $n$ for different sets of the
other parameters including the sum $n_+=(n_1+n)$ . In
each case, the energy at $n=0$, i.e., in the absence of
the second component, was subtracted from the total
energy. Strictly speaking, sequences of states with a
fixed chemical potential of the first background component 
are of interest. This condition is practically close to 
the equality  $(n_1+n)=$ const (with an accuracy of 
several percent). This accuracy is quite sufficient for
the qualitative analysis.

Most of the plots shown in Fig. 1 have a nontrivial minimum
of the ratio $\varepsilon(n)/n$. If $\varepsilon(n)$ is considered
as the equation of state of a certain fictitious medium
relating its particle number density and the internal
energy density, this minimum corresponds to zero pressure. 
The pressure at slightly higher and lower densities, is positive 
and negative (i.e., tension), respectively. Consequently, 
states homogeneous in the variable $\tau$ in a certain adjacent
$n$ region are stable. However, if ``cleavages'' are made in 
a medium with tension, it is collapsed into one-dimensional drops
adjacent to ``vacuum.'' In our three-dimensional system, 
each such drop is an elongated bubble inside the
first component filled with the second component.
Two steady-state two-dimensional configurations are
joined at each of two ends of the bubble; one of them is ``vacuum,''
i.e., the first component with separated empty vortices 
and the second configuration is the ``medium,'' i.e., 
one multiple vortex with the filled core.

\begin{figure}
\begin{center} 
\epsfig{file=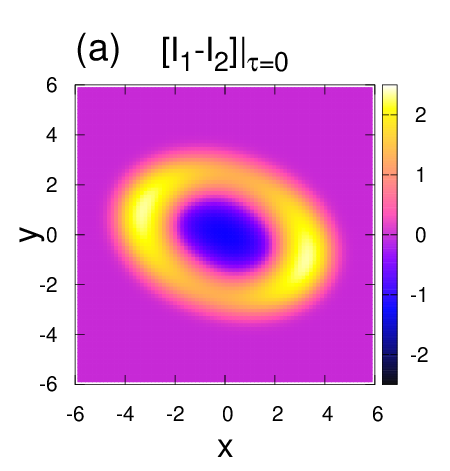, width=42mm}
\epsfig{file=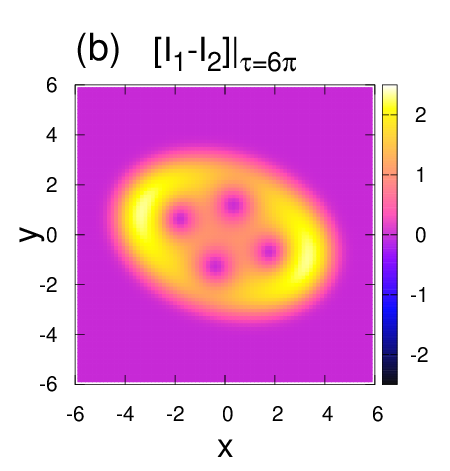, width=42mm}
\end{center}
\caption{
Cross sections of the light beam for the bubble with four vortices at
$\Omega=0.5$, $\langle n_+\rangle =60$, and $\zeta=590$.
}
\label{V4B} 
\end{figure}

\begin{figure}
\begin{center} 
\epsfig{file=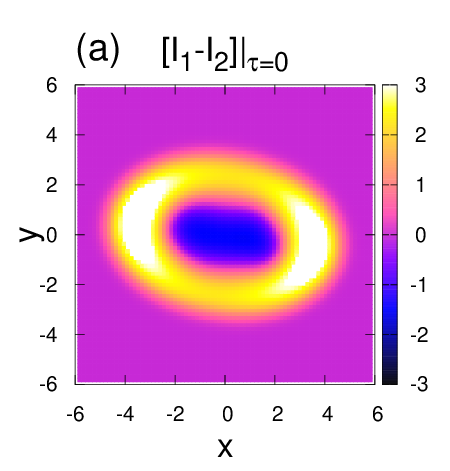, width=42mm}
\epsfig{file=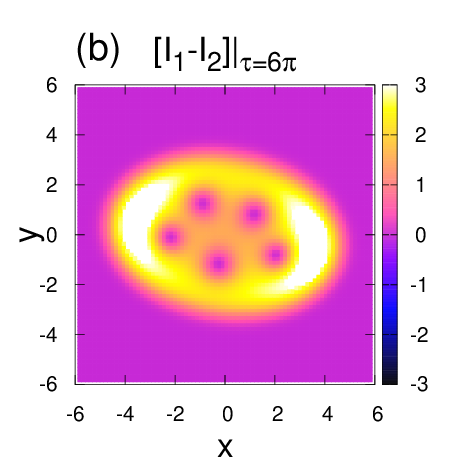, width=42mm}
\end{center}
\caption{
Cross sections of the light beam for the bubble with five vortices at
$\Omega=0.6$, $\langle n_+\rangle =100$, and $\zeta=560$.
}
\label{V5B} 
\end{figure}

\begin{figure}
\begin{center} 
\epsfig{file=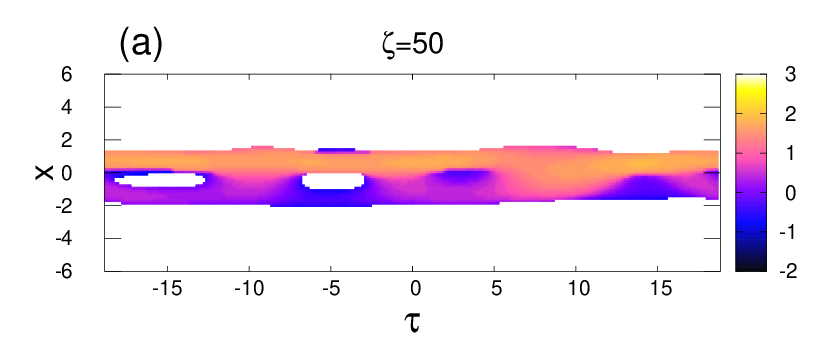, width=86mm}\\
\epsfig{file=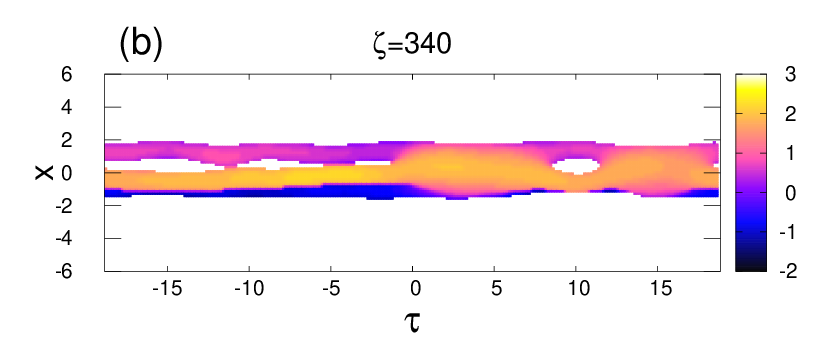, width=86mm}\\
\epsfig{file=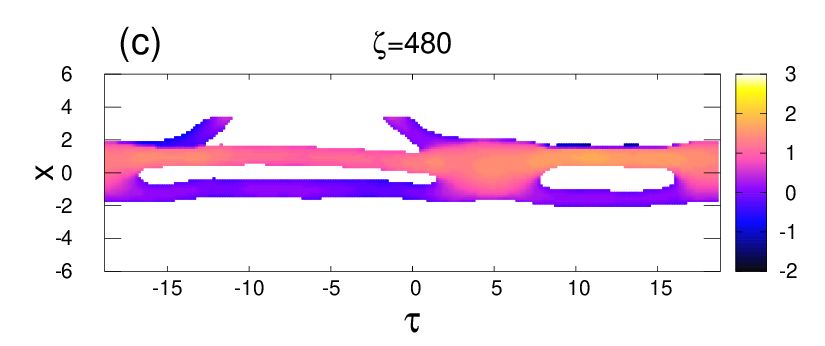, width=86mm}\\
\epsfig{file=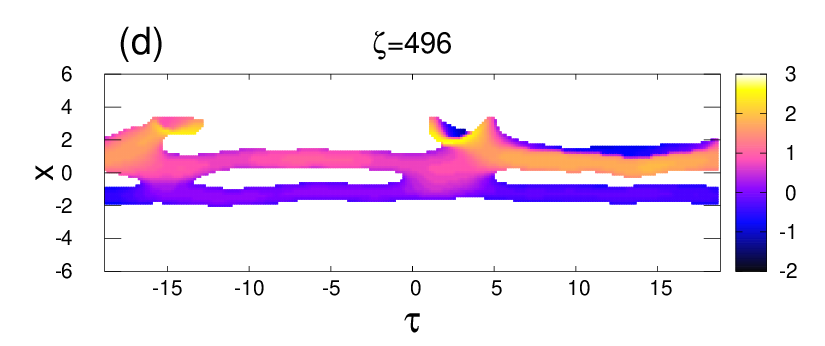, width=86mm}
\end{center}
\caption{
Example of the evolution of the slightly filled triple vortex at
the parameters $\Omega=0.4$ and $\langle n_+\rangle =60$
for which the corresponding two-dimensional system does not
have a steady-state solution with $n=0$ in the main stages of 
(a) the beginning of the formation of bubbles; 
(b) the appearance of a sufficiently long segment
with unfilled vortex filaments; (c) the reconnection of one
of the filaments with the waveguide wall (the wall is not
shown; for this reason, the filament looks as broken); and
(d) the output of the second component from the bubble to
the surface of the light beam.
}
\label{V3_04_60} 
\end{figure}

The direct numerical simulation with appropriate initial 
states confirmed the existence and stability of these structures. 
The first two examples (for bubbles with two and three vortices)
are presented in Fig. 2 and 3, where panels (a--c) demonstrate 
color maps of the difference between the intensities of the 
first and second wave components in three different sections 
of the light beam. Since the phase separation regime occurs
(i.e., the product $I_1 I_2$ is negligibly small almost always
except for a relatively narrow transient region of the
domain wall), regions with a positive difference (bright
colors) are filled almost exclusively with the first component,
whereas regions with a negative difference (dark colors) are 
filled primarily with the second component. Therefore, 
it is not necessary to represent each intensity field 
individually in figures. The difference $I_1-I_2$ does not
provide complete information only in the domain wall but this
information is not too important for the general understanding
of the picture. Figures 2d and 3d show the side views of the
domain wall, which is the boundary of the bubble, and
vortex filaments attached to it. Pictures for bubbles
with four and five vortices are presented in Figs. 4 and 5. 
Since the structures of all bubbles are qualitatively the same, 
only cross sections are given in the figures. It is noteworthy
that the cross section of bubbles at a large number of vortices
is ``oblate'' and empty vortices are located primarily along an
ellipse rather than a circle.

States in all above examples were not strictly steady
because small perturbations still remain after the
preparational dissipative procedure, but they did not
grow at long propagation distances, thus confirming
the stable character of the found structures.

It is worth noting that the two-dimensional system
of empty vortices is unstable at some parameter values:
the width of the waveguide is insufficient to contain
vortices, whereas the filled vortex of the corresponding
multiplicity is more compact and exists in a wide $n$
range (e.g., at $\Omega=0.4$, $Q=3$, $n_+=60$, and that is why 
the corresponding plot is absent in Fig. 1a). In these cases,
the ``underfilled'' three-dimensional vortex first evolves to
the formation of a bubble, and then, if deserted segments 
of vortex filaments are long enough, they gradually approach 
the waveguide wall and are recoupled to it. The second component 
leaves through the formed channels to the beam surface. 
Finally, the entire structure is destroyed. The corresponding
example is presented in Fig. 6. Another possible scenario 
of the development occurs when the bubble itself becomes locally
unstable with respect to transverse displacement and is discharged
to the waveguide wall (is not shown).

\subsection*{Conclusions}

To summarize, the numerical simulation has shown that rotation 
generated by the helical waveguide strongly changes the dynamics 
of the optical wave. In particular, it makes the existence of 
new-type structures theoretically possible. 
These structures significantly supplement the ``collection'' 
of known three-dimensional solitons [2]. Although these results
are obtained within the fundamental model given by Eq. (1) 
they are yet abstract at the current stage. A further work both
in theory in order to obtain analytical estimates for 
$\varepsilon(n)$ etc. and in the selection of materials
with the necessary properties for a possible future
experiment is necessary. Serious technical difficulties
in the way toward the implementation of such an
experiment can hardly be foreseen.

\subsection*{Funding}

This work was supported by the Ministry of Science and
Higher Education of the Russian Federation (state 
assignment no. FFWR-2024-0013).

\subsection*{Conflict of interests}

The author of this work declares that he has no conflicts of interest.


\begin{thebibliography}{99}

\bibitem{opt1}  Y. Kivshar and G. P. Agrawal,
{\it Optical Solitons: From Fibers to Photonic Crystals} 
(Academic, CA, 2003).

\bibitem{opt2} B. A. Malomed, {\it Multidimensional Solitons}, 
(AIP, Melville, 2022).\\
\url{https://doi.org/10.1063/9780735425118}

\bibitem{opt3} F. Baronio, S. Wabnitz, and Yu. Kodama,
Phys. Rev. Lett. {\bf 116}, 173901 (2016).

\bibitem{defocNLSE} P. G. Kevrekidis, D. J. Frantzeskakis,
and  R. Carretero-Gonz\'alez,
{\it The  Defocusing  Nonlinear  Schr\"odinger  Equation:
From Dark Solitons to Vortices and Vortex Rings} (SIAM, Philadelphia, 2015).

\bibitem{disp_manag} V. N. Serkin and A. Hasegawa, 
JETP Lett. {\bf 72}, 89 (2000).

\bibitem{dissip-1} S. K. Turitsyn, N. N. Rozanov, I. A. Yarutkina, 
A. E. Bednyakova, S. V. Fedorov, O. V. Shtyrina, 
and M. P. Fedoruk, Phys. Usp. {\bf 59}, 642 (2016).

\bibitem{dissip-2} N. A. Veretenov, N. N. Rozanov, and S. V. Fedorov,
Phys. Usp. {\bf 65}, 131 (2022).

\bibitem{LB_RA2000} S. Raghavan and G. P. Agrawal, Opt. Commun. {\bf 180}, 377 (2000). 

\bibitem{LB_JOB2005} B. A. Malomed, D. Mihalache, F. Wise, and L. Torner,
J. Opt. B {\bf 7}, R53 (2005).

\bibitem{LB_RW2013} W. H. Renninger and F. W. Wise,
Nat. Commun. {\bf 4}, 1719 (2013).

\bibitem{LB_PhysRevX2013}
F. Eilenberger, K. Prater, S. Minardi, R. Geiss, U. R\"opke, J. Kobelke, 
K. Schuster, H. Bartelt, S. Nolte, A. T\"unnermann, and T. Pertsch,
Phys. Rev. X {\bf 3}, 041031 (2013).

\bibitem{LB_SFKT2018} O. V. Shtyrina, M. P. Fedoruk, Y. S. Kivshar, and S. K. Turitsyn,
Phys. Rev. A {\bf 97}, 013841 (2018).

\bibitem{LB-2color} S. V. Sazonov, A. A. Kalinovich, M. V. Komissarova, and I. G. Zakharova,
Phys. Rev. A {\bf 100}, 033835 (2019).

\bibitem{LB_JETPL2021} E. D. Zaloznaya, A. E. Dormidonov, V. O. Kompanets,
S. V. Chekalin, and V. P. Kandidov, JETP Lett. {\bf 113}, 787 (2021).

\bibitem{LB_OptComm2023} P. Parra-Rivas, Y. Sun, and S. Wabnitz,
Opt. Comm. {\bf 546}, 129749 (2023).

\bibitem{R2024-2} V. P. Ruban, JETP Lett. {\bf 119}, 585 (2024).

\bibitem{BZ1970} A. L. Berkhoer and V. E. Zakharov, 
Sov. Phys. JETP {\bf 31}, 486 (1970).

\bibitem{mix1} Tin-Lun Ho and V. B. Shenoy, Phys. Rev. Lett. {\bf 77}, 3276 (1996).

\bibitem{mix2} H. Pu and N. P. Bigelow, Phys. Rev. Lett. {\bf 80}, 1130 (1998).

\bibitem{mix3} B. P. Anderson, P. C. Haljan, C. E. Wieman, and E. A. Cornell,
Phys. Rev. Lett. {\bf 85}, 2857 (2000).

\bibitem{mix4} S. Coen and M. Haelterman, Phys. Rev. Lett. {\bf 87}, 140401 (2001).

\bibitem{mix5} G. Modugno, M. Modugno, F. Riboli, G. Roati, and M. Inguscio, 
Phys. Rev. Lett. {\bf 89}, 190404 (2002).

\bibitem{sep} E. Timmermans, Phys. Rev. Lett. {\bf 81}, 5718 (1998).

\bibitem{AC1998} P. Ao and S. T. Chui, Phys. Rev. A {\bf 58}, 4836 (1998).

\bibitem{PDW1} M. Haelterman and A. P. Sheppard, Phys. Rev. E {\bf 49}, 3389 (1994).

\bibitem{PDW2} M. Haelterman and A. P. Sheppard, Phys. Rev. E {\bf 49}, 4512 (1994).

\bibitem{PDW3} A. P. Sheppard and M. Haelterman, Opt. Lett. {\bf 19}, 859 (1994).

\bibitem{PDW4} Yu. S. Kivhsar and B. Luther-Davies, Phys. Rep. {\bf 298}, 81 (1998).

\bibitem{PDW5} N. Dror, B. A. Malomed, and J. Zeng,
Phys. Rev. E {\bf 84}, 046602 (2011).

\bibitem{OVS1} A. H. Carlsson, J. N. Malmberg, D. Anderson, M. Lisak, 
E. A. Ostrovskaya, T. J. Alexander, and Yu. S. Kivshar,
Opt. Lett. {\bf 25}, 660 (2000).

\bibitem{OVS2} A. S. Desyatnikov, L. Torner, and Yu. S. Kivshar,
Prog. Opt. {\bf 47}, 291 (2005).

\bibitem{R2021-1} V. P. Ruban, JETP Lett. {\bf 113}, 532 (2021).

\bibitem{R2021-3} V. P. Ruban, J. Exp. Theor. Phys. {\bf 133}, 779 (2021).

\bibitem{R2022-2} V. P. Ruban, JETP Lett. {\bf 116}, 329 (2022).

\bibitem{R2023-1} V. P. Ruban, JETP Lett. {\bf 117}, 292 (2023).

\bibitem{R2023-2} V. P. Ruban, JETP Lett. {\bf 117}, 583 (2023).

\bibitem{R2023-3} V. P. Ruban, J. Exp. Theor. Phys. {\bf 137}, 746 (2023).

\bibitem{anom-defoc-1} X. Liu, B. Zhou, H. Guo, and M. Bache,
Opt. Lett. {\bf 40}, 3798 (2015).

\bibitem{anom-defoc-2}
X. Liu and M. Bache, Opt. Lett. {\bf 40}, 4257 (2015).

\end{thebibliography}
\end{document}